\documentclass[aps,prl,reprint,showpacs,showkeys,preprintnumbers]{revtex4-1}

\usepackage[bookmarks=true]{hyperref}
\usepackage{graphicx,epsfig,color}
\usepackage[english]{babel}
\usepackage{amssymb,amsfonts,amsmath}
\usepackage{verbatim}
  \newcommand{\cF}{{\cal F}}

  \newcommand{\cN}{{\cal N}}

\newcommand{\be}{\begin{equation}} \newcommand{\ee}{\end{equation}}
\newcommand{\bea}{\begin{eqnarray}} \newcommand{\eea}{\end{eqnarray}}
\newcommand{\beann}{\begin{eqnarray*}}  \newcommand{\eeann}{\end{eqnarray*}}
\newcommand{\bfig}{\begin{figure}} \newcommand{\efig}{\end{figure}}
\newcommand{\ba}{\begin{array}} \newcommand{\ea}{\end{array}}
\newcommand{\bcen}{\begin{center}} \newcommand{\ecen}{\end{center}}
\newcommand{\btab}{\begin{tabular}} \newcommand{\etab}{\end{tabular}}

\newtheorem{Proposition}{Proposition}[section]

\newtheorem{Theorem}{Theorem}[section]
\newtheorem{Lemma}{Lemma}[section]
\newtheorem{Corollary}{Corollary}[section]

\newcommand{\bp}{\begin{Proposition}}   \newcommand{\ep}{\end{Proposition}}
\newcommand{\bt}{\begin{Theorem}}   \newcommand{\et}{\end{Theorem}}
\newcommand{\bl}{\begin{Lemma}}     \newcommand{\el}{\end{Lemma}}
\newcommand{\bc}{\begin{Corollary}} \newcommand{\ec}{\end{Corollary}}
\begin{document}

%%%%%%%%%%%%%%%%%%%%%%%%%%%%%%%%%%%%%%
%%%%%%%%%%%%% TITLEPAGE %%%%%%%%%%%%%%
%%%%%%%%%%%%%%%%%%%%%%%%%%%%%%%%%%%%%%

\title{ Holographic quark matter and neutron stars}

\author{Carlos Hoyos}
\email{hoyoscarlos@uniovi.es}
\affiliation{Department of Physics, Universidad de Oviedo\\
Avda.~Calvo Sotelo 18, ES-33007 Oviedo, Spain}

\author{Niko Jokela}
\email{niko.jokela@helsinki.fi}
\affiliation{Department of Physics and Helsinki Institute of Physics\\
P.O.~Box 64, FI-00014 University of Helsinki, Finland}

\author{David \surname{Rodr\'{\i}guez Fern\'andez}}
\email{rodriguezferdavid@uniovi.es}
\affiliation{Department of Physics, Universidad de Oviedo\\
Avda.~Calvo Sotelo 18, ES-33007 Oviedo, Spain}

\author{Aleksi Vuorinen}
\email{aleksi.vuorinen@helsinki.fi}
\affiliation{Department of Physics and Helsinki Institute of Physics\\
P.O.~Box 64, FI-00014 University of Helsinki, Finland}

\begin{abstract}
We use a top-down holographic model for strongly interacting quark matter to study the properties of neutron stars. When the corresponding Equation of State (EoS) is matched with state-of-the-art results for dense nuclear matter, we consistently observe a first order phase transition at densities between two and seven times the nuclear saturation density. Solving the Tolman-Oppenheimer-Volkov equations with the resulting hybrid EoSs, we find maximal stellar masses in the excess of two solar masses, albeit somewhat smaller than those obtained with simple extrapolations of the nuclear matter EoSs. Our calculation predicts that no quark matter exists inside neutron stars.
\end{abstract}

\preprint{FPAUO-16/06}
\preprint{HIP-2016-08/TH}


\maketitle

\section{Introduction}

Quantitatively predicting the thermodynamic properties of dense nuclear and quark matter is one of the main challenges of modern nuclear theory. The complexity of the task originates from the need to nonperturbatively solve the theory of strong interactions, QCD, at finite baryon chemical potential $\mu_B$. This combination of requirements is problematic, as it makes all the usual first principles tools  fail: Lattice simulations suffer from the infamous sign problem at finite baryon chemical potential \cite{Cristoforetti:2012su}, while perturbative QCD is invalidated by the sizable value of the gauge coupling at moderate densities \cite{Kraemmer:2003gd}. At present, the Equation of State (EoS) of cold strongly interacting matter is under quantitative control at baryon densities below the nuclear saturation limit, $n_B\leq n_s\approx 0.16/\text{fm}^3$, where Chiral Effective Theory (CET) works \cite{Tews:2012fj,Gezerlis:2013ipa}, as well as at baryon chemical potential above roughly 2.5 GeV where the perturbative EoS converges \cite{Freedman:1976ub,Vuorinen:2003fs,Kurkela:2009gj,Kurkela:2016was}. These limits unfortunately exclude the densities $n_s\leq n_B \leq 10n_s$, where a deconfining phase transition to quark matter is expected to occur \cite{Fukushima:2013rx}. 

Remarkably, baryon densities well beyond the saturation limit are realized inside the most massive neutron stars \cite{Heiselberg:1999mq}. Due to the difficulties alluded to above, a microscopic description of these objects necessitates bold extrapolations of the CET results, typically relying on a systematic use of so-called polytropic EoSs \cite{Hebeler:2013nza}. The polytropic EoSs have as such no physical content, but simply parameterize our current ignorance of the high-density EoS in a way that allows constraining from both the low- and high-density sides \cite{Kurkela:2014vha}. The fact that no first principles results are available for ultradense nuclear matter or strongly coupled quark matter makes progress towards a quantitatively reliable neutron star matter EoS excruciatingly slow. 

Clearly, there is a need for fundamentally new approaches to the physics of strongly coupled quark matter --- a challenge not unlike understanding the dynamics of hot quark-gluon plasma \cite{Brambilla:2014jmp}. In this context, a very promising approach has turned out to be to apply the holographic duality \cite{Maldacena:1997re,Gubser:1998bc,Witten:1998qj}. It has been successfully used to study the deconfined phases of QCD matter \cite{Erdmenger:2007cm,Adams:2012th} and to probe very nontrivial equilibration dynamics \cite{Casalderrey-Solana:2013aba,Bantilan:2014sra,Chesler:2015wra}, teaching the heavy ion community many qualitative and even quantitative lessons about the behavior of strongly coupled QCD matter.

So far, holography has been used to study the cold and dense part of the QCD phase diagram only to a limited extent (see however \cite{Bergman:2007wp,Rozali:2007rx,Kim:2007vd,Kaplunovsky:2012gb,Li:2015uea}). The reason for this is that in its best understood limit, the duality deals with supersymmetric conformal field theories, which are fundamentally different than QCD. In particular, they typically contain only adjoint representation fields, and have therefore no analogue of the fundamental representation quarks that dominate the properties of cold and dense QCD matter.

Despite the above issues, the situation is not hopeless: In the 't Hooft limit of $\lambda_{YM}\equiv g_{YM}^2N_c\gg 1$ and $N_c\gg N_f$, the dynamics of fundamental flavors can be captured by degrees of freedom carried by probe D-branes, while the gluon sector continues to be described by classical supergravity (SUGRA) \cite{Karch:2002sh}. States with finite baryon density in the gauge theory correspond to gravity configurations with a gauge field turned on in the D-brane worldvolume. The free energy can then be computed by evaluating the classical on-shell action of SUGRA together with the D-brane action. Given the relative simplicity of the calculations involved, the duality thus bestows us with a powerful tool to explore strongly coupled quark matter even at high density.

Our goal in this paper is to take the logical step from the D3-D7 construction of \cite{Karch:2002sh} to phenomenological neutron star physics by investigating the implications of using a holographic EoS for cold quark matter just above the deconfinement transition. Due to technical restrictions discussed in the following section, completing this task requires some bold extrapolations. It will, however, lead us to results in excellent accordance with current phenomenological expectations, with only one parameter fitted to experiments. 

The paper is organized as follows: Our construction is thoroughly explained in \S2, while the resulting EoS and its relation to that of nuclear matter is analyzed in \S3. The implications of the hybrid EoS for the properties of neutron stars are then displayed in \S4, while conclusions are drawn and an outlook presented in \S5.

\section{Holographic model}

In order to describe quark matter at nonzero density, let us consider a D3-D7 brane intersection. The field theory is then $\cN=2$ Super Yang-Mills (SYM) with the matter content of $\cN=4$ $SU(N_c)$ SYM in the adjoint sector and $N_f$ matter hypermultiplets in the fundamental representation. Thus, in addition to the QCD quarks and gluons, there are squarks and several species of adjoint fermions and scalars. The theory has a global $U(N_f)\sim U(1)_B\times SU(N_f)$ flavor symmetry,  the $U(1)_B$ part of which we identify as the baryon symmetry. For two flavors, i.e.~$N_f=2$, isospin is the Abelian subgroup $U(1)_I\subset SU(2)$. Note that both quarks and squarks are charged under the flavor symmetry, so a typical state will have a finite density of both types of particles. 
Also, we do not expect our model to capture the correct gluon dynamics, as it has exact superconformal invariance.

In the large-$N_c$ limit and at strong 't Hooft coupling, the $\cN=4$ SYM theory has a holographic description in terms of classical type IIB SUGRA in an $AdS_5\times S^5$ geometry \cite{Maldacena:1997re}. In the 't Hooft limit $N_f\ll N_c$, the flavor sector can be introduced as $N_f$ probe D7-branes extended along the $AdS_5$ directions and wrapping an $S^3\subset S^5$ \cite{Karch:2002sh}. 
The thermodynamic properties of the model have been studied in great detail at nonzero temperature and charge density \cite{Mateos:2006nu,Kobayashi:2006sb,Mateos:2007vn,Karch:2007br,Nakamura:2007nx,Ghoroku:2007re,Mateos:2007vc,Erdmenger:2008yj,Ammon:2008fc,Basu:2008bh,Faulkner:2008hm,Ammon:2009fe,Erdmenger:2011hp}. The free energy can be split in the contributions of adjoint and flavor fields
\begin{equation}
F=F_{\cN=4}+F_{\text{flavor}},
\end{equation}
where the first term is independent of the charge density and does not play a very important role for us. 

We work in the grand canonical ensemble, so that the free energy is a function of the temperature $T$ as well as chemical potentials corresponding to the conserved charges. Barring the presence of a mixture of two phases, possible in a first order transition, the matter inside neutron stars is typically taken to be locally charge neutral and in beta equilibrium. This can be realized by taking the chemical potentials and densities of the $u$, $d$, and $s$ quarks to agree \cite{1996csnp.book.....G}, which implies neglecting the differences in their bare masses and setting both the isospin chemical potential and electron density to zero. In the zero-temperature limit, relevant for quiescent neutron stars, the EoS can then be parameterized by the baryon chemical potential $\mu_B=N_c \mu_q$ alone. In this case the holographic setup simplifies somewhat, as there is no spontaneous breaking of flavor symmetry in the 't Hooft limit \cite{Erdmenger:2008yj,Ammon:2008fc,Basu:2008bh,Faulkner:2008hm,Ammon:2009fe}. 

In the limit explained above, the flavor contribution to the grand canonical free energy density reads \cite{Karch:2007br,Karch:2008fa,Karch:2009eb,Ammon:2012je,Itsios:2016ffv}
\begin{equation}\label{eq:freen}
\cF_{\text{flavor}}=-\frac{N_c N_f}{4 \gamma^3 \lambda_{YM}}(\mu_q^2-m^2)^2+{\mathcal O}\left(\mu_q^3\, T,T^4\right),
\end{equation}
where $\gamma\equiv\Gamma(7/6)\Gamma(1/3)/\sqrt{\pi}$ and $m$ is a mass parameter associated with the fermions. The model has thus four parameters: The number of colors $N_c$, the number of flavors $N_f$, the 't Hooft coupling $\lambda_{YM}$, and the mass $m$ appearing in the dimensionless ratio $\mu_q/m$. We choose them according to the properties of deconfined QCD matter at the relevant densities, which implies setting $N_c=N_f=3$. The contribution of the adjoint sector to the free energy $\cF_{\cN=4}\sim N_c^2 T^4$ becomes of the same order as the ${\mathcal O}(T^4)$ corrections to the flavor free energy, and can thereby be neglected.

Upon choosing the above values for $N_c$ and $N_f$, we are extrapolating our model to a regime where finite $N_c$ and $N_f/N_c$ corrections are expected to become important \cite{Gubser:1998nz,Bigazzi:2009bk,Nunez:2010sf,Bigazzi:2011it,Bigazzi:2013jqa}. For practical reasons, we however neglect them in the following, which implies that we treat the model as phenomenologically motivated by the original string theory construction. We also allow $\lambda_{YM}$ and $m$ to take values appropriate for the physical system under consideration, expecting them to lie in a region where the holographic approach remains at least qualitatively valid (for a recent discussion of the convergence of strong coupling expansions, see \cite{Waeber:2015oka}). 

%%%%%%%%%%%%%%%%%%%%%%%%%%%%%%%%%%%%%%%%%%%%%%%%%%%%%%
\begin{figure*}[ht]
\center
\includegraphics[width=0.44\textwidth]{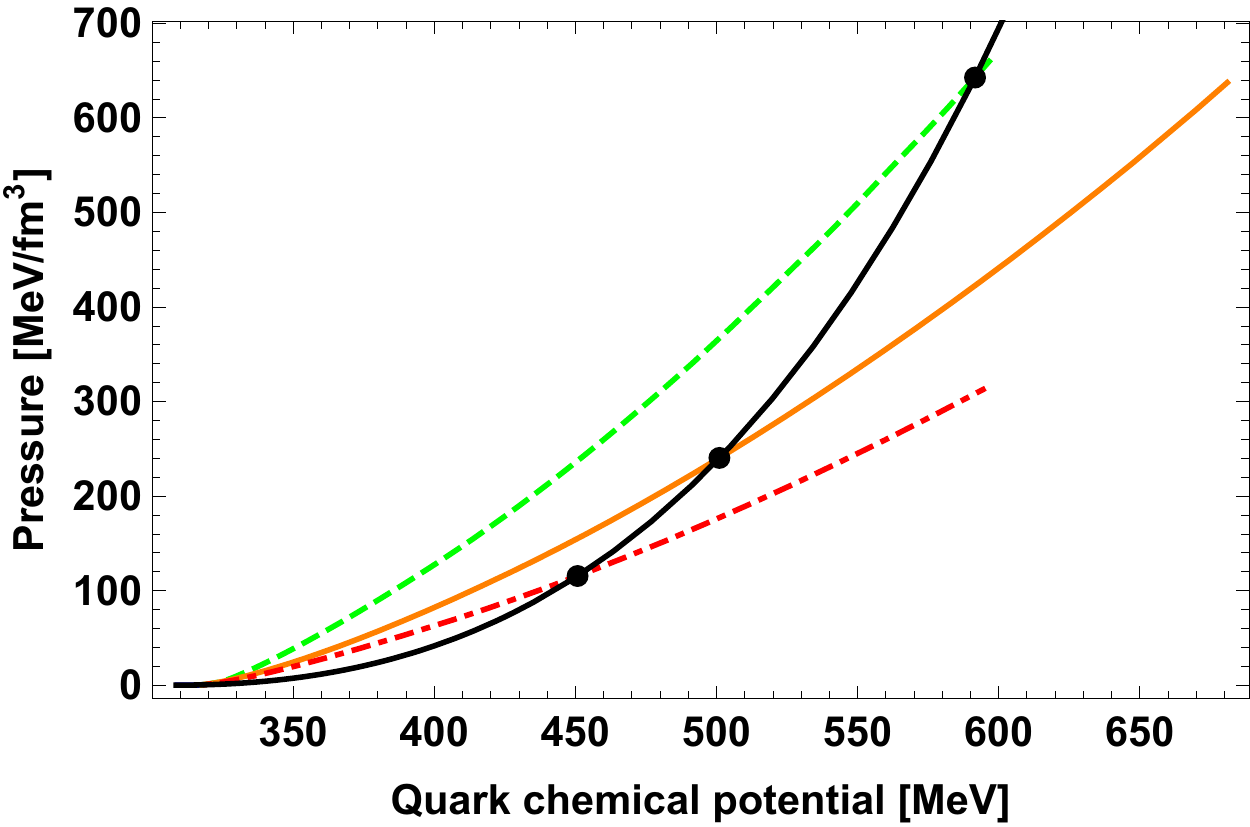}
$\;\;\;\;\;$ \includegraphics[width=0.45\textwidth]{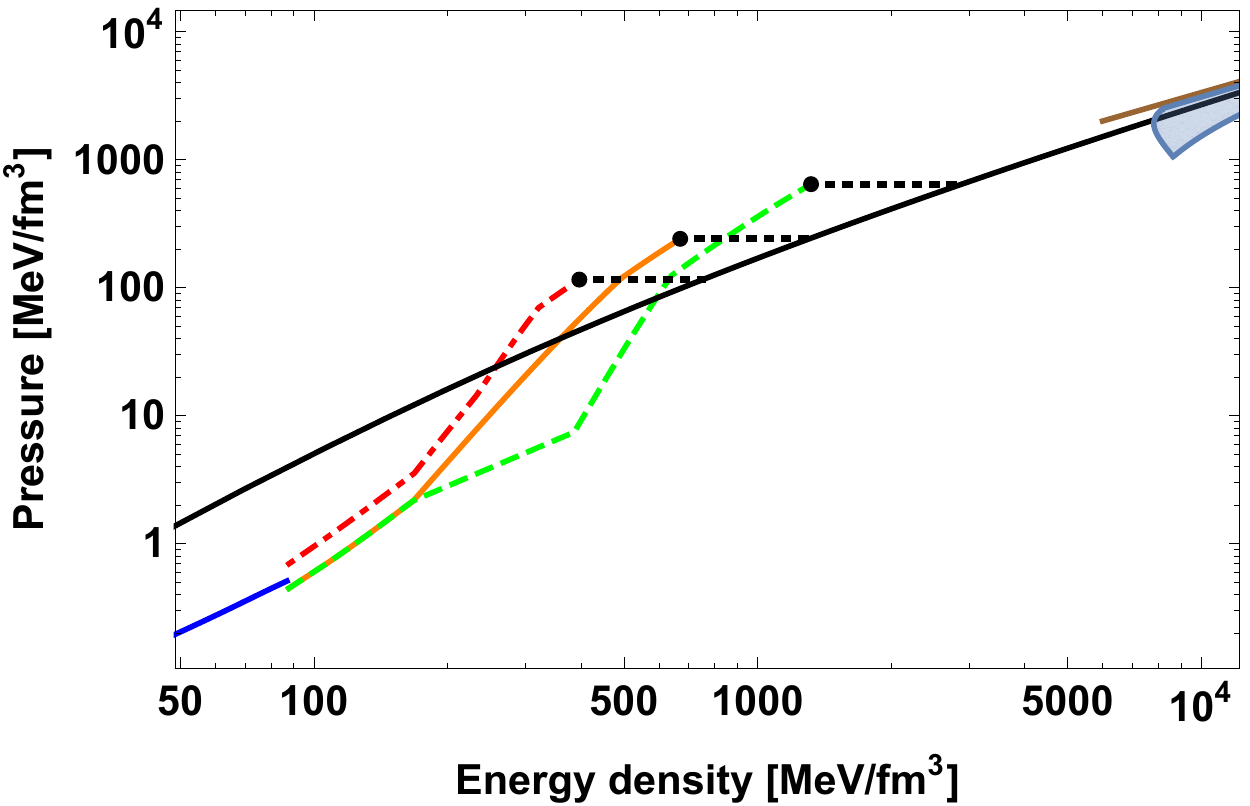}
\caption{Left: The holographic quark matter EoS (black curve) together with the nuclear matter EoSs of \cite{Hebeler:2013nza}: Soft (green), intermediate (orange), and stiff (red). Right: The matching procedure from the low-energy EoSs to the quark matter one, with the dashed black lines showing the jump in the energy density, characteristic of a first order transition. Shown are also the CET results of \cite{Tews:2012fj,Gezerlis:2013ipa} (blue curve), the conformal limit (brown curve), and the perturbative result of \cite{Kurkela:2009gj} (light blue band, generated by varying the renormalization scale).}
 \label{fig:PvsmuE}
\end{figure*}
%%%%%%%%%%%%%%%%%%%%%%%%%%%%%%%%%%%%%%%%%%%%%%%%%%%%%%

With the above reservations, we proceed to note that in the limit of large chemical potentials, the free energy density of our model approaches the value
\begin{equation}
\cF_{\text{flavor}}\to -\frac{N_c N_f}{4 \gamma^3 \lambda_{YM}}\mu_q^4,
\end{equation}
the form of which is fixed by conformal invariance in the UV.  In QCD, the corresponding quantity is known to approach the Stefan-Boltzmann value \cite{Freedman:1976ub}
\begin{equation}
\cF_{QCD}\to -\frac{N_c N_f}{12\pi^2}\mu_q^4,
\end{equation}
so imposing the requirement that our model has the correct limiting behavior at large density fixes the value of the 't Hooft coupling as $\lambda_{YM}=3\pi^2/\gamma^3\simeq 10.74$.
With this choice, our model can be seen to match the perturbative EoS of \cite{Kurkela:2009gj} already at moderate densities. 
 
Finally, we discuss the choice of the mass parameter $m$. We expect that in the strongly coupled region the effective masses of the quarks receive large nonperturbative corrections, so relating this last remaining parameter of our model to the (differing) bare masses of the $u$, $d$, and $s$ quarks would be largely nonsensical. Rather, we fix $m$ through the value of $\mu_q$, where the pressure of our model vanishes, requiring it to agree with the value obtained from the EoS of nuclear matter \cite{Negele:1971vb}. This gives $m \approx 308.55\, \text{MeV}$, just below one third of the nucleon mass.

As argued above, at large densities and vanishing temperature, the pressure $p$ and the energy density $\varepsilon$ of our model can be determined from Eq.~\eqref{eq:freen} as $p=-\cF_{\text{flavor}}$, $\varepsilon=\mu_q\frac{\partial p}{\partial \mu_q}-p$. The EoS thus takes the simple form
\begin{eqnarray}\label{eqeos}
\!\!\varepsilon\,=\,3p+m^2\sqrt{\frac{N_c N_f}{4\gamma^3 \lambda_{YM}}p} 
\,=\, 3p+\frac{\sqrt{3}m^2}{2\pi}\sqrt{p},
\end{eqnarray}
while the speed of sound squared reads $c_s^2\,=\, \frac{\partial p}{\partial \varepsilon}$.
From \eqref{eqeos}, $c_s^2$ always resides below the conformal value of $1/3$, making our EoS comparatively soft, seemingly at odds with the conclusions of Ref.~\cite{Bedaque:2014sqa}. It should, however, be noted that in \cite{Bedaque:2014sqa} the transition between the nuclear and quark matter phases was fixed to occur at twice the nuclear saturation density. In our case, this parameter is one of the predictions of the model, and its value turns out to be always somewhat larger than $2n_s$.

\section{Matching to nuclear matter}
 
Having obtained a candidate EoS for strongly coupled dense quark matter, the natural question arises, how to best use it in applications within neutron star physics. At low densities, we expect the matter to reside in the confined phase and, as the density is increased, find a transition to deconfined matter. This transition cannot be realized  purely within the D3-D7 model,  because at nonzero baryon density quarks are always in a deconfined phase, at least in the large-$N_c$ limit \footnote{This is not the case for other holographic models, such as the Sakai-Sugimoto one \cite{Sakai:2004cn}; cf.~\cite{Kim:2006gp,Bergman:2007wp,Rozali:2007rx,Kim:2007vd,Rho:2009ym,Kaplunovsky:2012gb,deBoer:2012ij,Kaplunovsky:2013iza,Kaplunovsky:2015zsa,Li:2015uea}. Attempts in this direction have, however, led to either unstable or unrealistic stars \cite{Burikham:2010sw,Kim:2011da,Ghoroku:2013gja,Kim:2014pva}.}. The most natural strategy is therefore to describe the low-density phase using state-of-the-art results from the CET of nuclear interactions below saturation density, extrapolated to higher densities with polytropic EoSs \cite{Hebeler:2013nza}. We then compare the corresponding pressure, i.e.~minus the free energy density, to that of our holographic system, thereby determining the dominant phase at each quark chemical potential. Due to the uncertainty related to the low-density result, the matching should not be performed using a single confining EoS; instead, we apply the three EoSs given in Table 5 of \cite{Hebeler:2013nza}, dubbed `soft', `intermediate', and `stiff', to represent different possible behaviors of the nuclear matter EoS. Of the three, the soft and stiff EoSs correspond to extreme cases, while the intermediate one can be considered a typical low-density EoS.

Our detailed construction is shown in Fig.~\ref{fig:PvsmuE}, where on the left side we display the three low-density EoSs together with our quark matter EoS in the form of pressure vs.~quark chemical potential. As can be seen from here, there is a critical chemical potential $\mu_{crit}$ for each of the three low-density EoSs, at which a phase transition to deconfined quark matter occurs. In all cases, the transition is of first order, which can be verified from the right figure that displays the hybrid EoSs on a logarithmic pressure vs.~energy density plane. Notice that the holographic quark matter EoS smoothly connects to the perturbative one of \cite{Kurkela:2009gj} at high density.

It is interesting to note that the densities, at which the first order phase transitions occur, are consistently in a phenomenologically viable region: For the soft nuclear matter EoS we get $n_{crit}=6.92\, n_s$, for the intermediate one $n_{crit}=3.79\, n_s$, and for the stiff case $n_{crit}=2.37\, n_s$. This strengthens our conclusion that the holographic description is consistent with the expected properties of strongly coupled quark matter at least on a qualitative level.  The order of the transition is, however, highly sensitive to the details of the EoS near the transition, and may therefore be smoother than we predict.

\section{Neutron star structure}

The EoS of strongly interacting matter is in a one-to-one correspondence with the Mass-Radius relation of neutron stars. This link is provided by the Tolman-Oppenheimer-Volkov (TOV) equations that govern hydrostatic equilibrium inside the stars. The equations take as input the relation between the energy density $\varepsilon$ and pressure $P$ of the matter, i.e.~its EoS, as well as the central energy density $\varepsilon(r=0)$, and produce the mass and radius of the corresponding star. Varying $\varepsilon(r=0)$, we then obtain a well-defined curve on the $MR$-plane.

A subtlety related to systems where a first order phase transition occurs is the possible existence of mixed phases. This, however, strongly depends on the value of the microscopic surface tension between the nuclear and quark matter phases. As this parameter is beyond the validity of our description, and only crude estimates for the quantity exist in QCD, we have chosen to neglect this scenario and only consider stars made of pure phases.

\begin{figure}[ht]
\center
\includegraphics[width=0.45\textwidth]{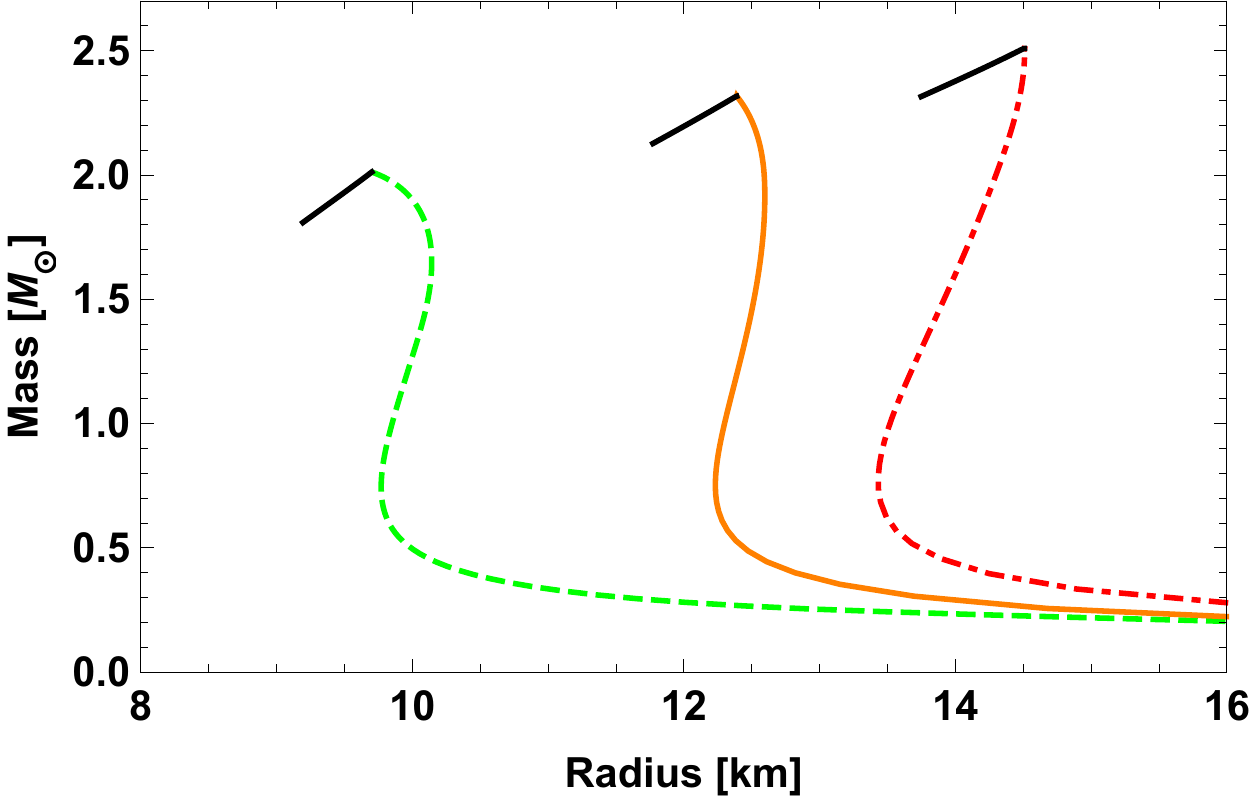}
\caption{The Mass-Radius relations corresponding to the three matched EoSs of Fig.~\ref{fig:PvsmuE} (right). The black lines correspond to an unstable branch of stars containing quark matter. The forms of the $M$-$R$ relations are fairly generic, see e.g.~\cite{1996csnp.book.....G}.}
 \label{fig:PMvsR}
\end{figure}

Plugging the three EoSs of Fig.~\ref{fig:PvsmuE} into the TOV equations, we obtain the Mass-Radius curves displayed in Fig.~\ref{fig:PMvsR}. They follow the corresponding curves of Ref.~\cite{Hebeler:2013nza} until they abruptly come to an end at points that mark the densities of our first order phase transition. Here, the solutions to the TOV equation take a sharp turn towards smaller masses and radii, signaling an instability with respect to radial oscillations \cite{1996csnp.book.....G}. This behavior follows from the sizable latent heat $\Delta Q=\mu_{crit}\Delta n$ at our first order transition, i.e.~the fact that the transitions are relatively strong for all three nuclear matter EoSs due to the softness of the holographic EoS (cf.~\cite{Kurkela:2014vha} and fig.~6 therein). The values we find for $\Delta Q$ are $(331\, \text{MeV})^4$ (soft),   $(265\, \text{MeV})^4$ (intermediate), and $(229\, \text{MeV})^4$ (stiff). 

%%%%%%%%%%%%%%%%%%%%%%%%%%%%%%%%%%%%%%%%%%%%%%%%%%%
%%%%%%%%%%%%%%%%%%%%%%%%%%%%%%%%%%%%%%%%%%%%%%%%%%%

The main conclusion to be drawn from our results is that with quark matter following a holographic EoS, it is unlikely that any deconfined matter could be found inside neutron stars. The maximal masses of the stars are dictated by the densities at which a phase transition from nuclear to quark matter occurs, with the most massive star having a central density at exactly this value. For the three nuclear matter EoSs of \cite{Hebeler:2013nza}, we find maximal masses of 2.01, 2.32, and 2.50 times the solar mass $M_{\odot}$, corresponding to radii of 9.7, 12.4, and 14.5km.

\section{Conclusions and outlook}

Neutron stars provide a unique laboratory for the study of cold ultradense nuclear matter --- and possibly even deconfined quark matter. 
Recent years have witnessed remarkable progress in their observational study, with the detection of the first two solar mass star already ruling out several models of dense nuclear matter \cite{Demorest:2010bx} and the recent discovery of gravitational waves by the LIGO and Virgo collaborations raising hopes of a dramatic improvement in the accuracy of radius measurements \cite{Abbott:2016blz}. This poses a prominent challenge for the theory community, and highlights the need to understand the properties of dense nuclear and quark matter from first principles.

In this paper, we have taken first steps towards the goal of building a phenomenological description for real world quark matter using holography. 
Under the usual large-$N_c$ and strong coupling assumptions it is possible to find a simple analytic expression for the EoS, which we, however, need to extrapolate to a regime where sizable corrections are to be expected. An important additional caveat is that the phase diagram of the theory may possess nontrivial structure; for instance, it was argued in \cite{Chen:2009kx} that at low temperatures squarks may condense and the system resides in a Higgsed phase. No other instabilities have been found \cite{Bigazzi:2013jqa}, but the appearance of spatially modulated phases  is not ruled out \cite{Domokos:2007kt,Nakamura:2009tf,Donos:2011bh,Bergman:2011rf}.

Despite the above limitations, the predictions of our model display remarkably good agreement with those of complementary approaches (see e.g.~\cite{Hebeler:2013nza,Kurkela:2014vha} and references therein). After fixing the parameters of our setup in a simple way, we obtained results that consistently indicate the presence of a strong first order deconfinement transition between the nuclear and quark matter phases at baryon densities between roughly two and seven times the nuclear saturation density. Due to the sizable latent heat associated with the transition, we predict that no stars with quark matter cores exist: As soon as there is even a small amount of quark matter in the center of a neutron star, it becomes unstable with respect to radial oscillations.

There exist a number of directions, in which our current work can be generalized. The obvious extension would be to allow a mixed phase of nuclear and quark matter, assuming a given value for the surface tension between the two phases \cite{Palhares:2010be}. In addition, one may  consider corrections due to the different bare masses of the quark flavors, as well as to nonzero temperature or background magnetic fields. With moderate effort, one may also consider the effects of finite $N_c$ and $\lambda_{YM}$ corrections on the EoS, utilizing existing results at the Next-to-Leading Order level.  Finally, an important strength of holography lies of course in its applicability to the determination of quantities that are very challenging for traditional field theory techniques. These include e.g.~transport constants and emission rates, which could both be considered within our present model.

An interesting, albeit also challenging direction to pursue would be to consider more refined top-down holographic models of QCD. One of the most appealing candidates is the-Sugimoto model \cite{Sakai:2004cn}, which has the same matter content as QCD at low energies and furthermore realizes confinement and chiral symmetry breaking in a natural way. As there are indications that this model exhibits a phase transition between baryonic and deconfined matter \cite{Li:2015uea}, it might enable performing the matching to the CET EoS at much lower densities where the uncertainty of the latter result is smaller. In the deconfined phase, the corresponding EoS is in addition significantly stiffer that that of a conformal theory \cite{Kulaxizi:2008jx,Jokela:2015aha}, which may lead to the existence of  stable stars with quark matter cores. A potential drawback of this approach is, however, that at very large densities it  deviates from QCD due to the lack of a UV fixed point.

%%%%%%%%%%%%%%%%%%%%%%%%%%%%%%%%%%%%%%%%%%%%
%\paragraph{Acknowledgments.-}
%%%%%%%%%%%%%%%%%%%%%%%%%%%%%%%%%%%%%%%%

\begin{acknowledgments}
{\em  Acknowledgments.-} We thank Aleksi Kurkela, Joonas N\"attil\"a, and Alfonso V. Ramallo for useful discussions. N.J.~and A.V.~have been supported by the Academy of Finland grants no.~273545 and 1268023, while C.H.~and D.R.F.~are partially supported by the Spanish grant MINECO-13-FPA2012-35043-C02-02. C.H.~is in addition supported by the Ramon y Cajal fellowship RYC-2012-10370, and D.R.F.~by the GRUPIN 14-108 research grant from Principado de Asturias.
\end{acknowledgments}

\bibliographystyle{apsrev4-1}

\bibliography{biblio}

\end{document}